\documentclass{llncs}
\usepackage{pst-node}
\usepackage{latexsym}
\usepackage{amsmath}
\usepackage{epsfig}
\usepackage{makeidx}  

\def\q5uad{\quad\quad\quad\quad\quad}

\begin{document}
\mainmatter              
\title{Malicious Software Indexing based on Behavioral Characteristics of System-call Dependency Graphs}

\title{Detecting Malicious Code based on System-call Dependency Graphs}
\title{Detecting Malicious Code based on Behavior Equivalence in System-call Dependency Graphs}
\title{Detecting Malicious Code based on System-call Dependency Graphs' Behavior Equivalence}
\title{Detecting Malicious Code based on Graphs \\ of System-call Group Dependencies}
\title{Detecting Malicious Code based on \\ Dependency Graphs of System-call Groups}

\title{Detecting Malicious Code by Exploiting Dependencies of System-call Groups}

\titlerunning{Detecting Malicious Code}  

\author{Stavros D. Nikolopoulos \ and \ Iosif Polenakis}
\authorrunning{Nikolopoulos and Polenakis} 
%
\tocauthor{Stavros D. Nikolopoulos and Iosif Polenakis}
\institute{Department of Computer Science \& Engineering, \\
University of Ioannina, GR-45110, Greece\\
\email{\{stavros,ipolenak\}@cs.uoi.gr}}

\maketitle              

\begin{abstract}
In this paper we present an elaborated graph-based algorithmic technique for efficient malware detection. More precisely, we utilize the system-call dependency graphs (or, for short ScD graphs), obtained by capturing taint analysis traces and a set of various similarity metrics in order to detect whether an unknown test sample is a malicious or a benign one. For the sake of generalization, we decide to empower our model against strong mutations by applying our detection technique on a weighted directed graph resulting from ScD graph after grouping disjoint subsets of its vertices. Additionally, we have developed a similarity metric, which we call NP-similarity, that combines qualitative, quantitative, and relational characteristics that are spread among the members of known malware families to archives a clear distinction between graph-representations of malware and the ones of benign software. Finally, we evaluate our detection model and compare our results against the results achieved by a variety of techniques proving the potentials of our model.
\end{abstract}


\section{Introduction}

A malicious software or  \emph{malware} may refer to any kind of software that its functionality is to cause harm to a user, computer, or network \cite{SiHo}. Thus, any software with malicious purposes can be considered as malware.
The most hard-to-detect malware mutation is the metamorphic malware. According to the definitions given in
\cite{YoKa}, {\it metamorphism} is the process of transforming a piece of code, utilizing a mutation module called {\it metamorphic engine}, responsible for the replication of malware into copies that are structurally different. However, these copies tend to exhibit the same behavior. Specifically, a very important clue upon which is based our detection approach, is the fact that every new copy has modified structure, code sequence size and syntactic properties \cite{RaMaSu}, while its behavior remains the same.

\vspace*{0.15in}
\noindent {\bf Malware Detection.} The term \textit{malware detection} is referred to the process of determining whether a given program $\pi$ is malicious or benign according to an a priori knowledge \cite{ChJhSeSoBr,MaHi,IdMa,AlLaVeWa}. For this purpose there have been proposed several techniques that leverage various characteristics for distinguishing malicious from benign programs. However, an efficient malware detection is based on an important process, called {\it malware analysis}, which collects the required information.

More precisely, malware analysis \cite{BaMoKrKi} is the process of determining the purpose and the functionality or, in general, the behavior of a given malicious code. Such a process is a necessary prerequisite in order to develop efficient and effective detection and also classification methods; malware analysis is divided into two main categories, namely {\it Static} and {\it Dynamic} analysis \cite{SiHo}.

\begin{itemize}
\renewcommand{\labelitemi}{\scriptsize$\circ$}
\item{\bf Static analysis:} In static analysis the specimen (i.e., test sample) is examined without its execution, performing the analysis on its source code.
\item{\bf Dynamic analysis:} In dynamic analysis an execution of the malware has to be performed in order to collect the required data, concerning the behavior of a program. However this approach needs more expertise while is extremely dangerous for the host environment. As a result, in most of times dynamic analysis is performed in a virtual environment.
\end{itemize}

It is well known that the behavior of a program can be modeled based upon system-call dependencies as they capture its interaction with its hosting environment, the operating system. As easily one can understand, a representation that captures a sequences of system-calls would be liable since any reorder or addition of one or more system-calls could change the sequence. Thus, a more flexible representation that would capture their in between relations, as a graph in example, could address that problem \cite{KoCoKrKiZhWa}.

As mentioned in \cite{FRJhChSaYa}, most malware relies on system-calls in order to deliver their payload. Additionally, since the behavior of a malware program could be reflected by the effect on its host operating system's state, then its behavior can be modeled by a directed acyclic graph, generated from system-call traces collected during its execution \cite{PaReMuSu}, the so called behavior graph.


\vspace*{0.15in}
\noindent {\bf Our Contribution.} In this paper we present an elaborated graph-based algorithmic technique that effectively addresses the problem of malware detection. Our approach for malware detection is based on the, so far unexploited, information that system-calls of a program $\pi$ of similar functionality can be classified into the same group and also on a set of various similarity metrics concerning the dependencies between these groups.

More precisely, having an instance of a ScD graph $D[\pi]$, constructed by the system-calls invoked by a program $\pi$, we decided to empower our model against strong mutations by applying our detection technique on a weighted directed graph $D^*[\pi]$, which we call group dependency graph (or, for short GrD graph), resulting from $D[\pi]$ after grouping disjoint subsets of its vertices. Additionally, we propose the similarity metric NP-similarity that combines similarity metrics on qualitative, quantitative, and relational characteristics that are spread among the members of known malware families to achieve a distinction between a malware and a benign program.

Finally, we evaluate our detection model and compare our results against the results achieved by a variety of techniques proving the potentials of our model.

\vspace*{0.15in}
\noindent {\bf Related Work.} Our model design is inspired by the use of system-call dependency graphs as described in \cite{BaReSo,ChJhKr,FRJhChSaYa}.


In~\cite{FRJhChSaYa}, Fredrikson {\it et al.} proposed an automatic technique for extracting optimally discriminative behavioral specifications, based on graph mining and concept analysis, that have a low false positive rate and at the same time are general enough, when used by a behavior based malware detector, to efficiently distinguish malicious from benign programs.

Christodorescu {\it et al.} \cite{ChJhKr} propose an algorithm that automatically constructs {\it specifications} of malicious behavior needed by AV's in order to detect malware. The proposed algorithm constructs such specifications by comparing the execution behavior of a known malware against the corresponding behaviors produced by benign programs.

Finally,  Babic {\it et al.} \cite{BaReSo} propose an approach to learn and generalize from the observed malware behaviors based on tree automate interference where the proposed algorithm infers $k$-testable tree automata from system-call data flow dependency graphs in order to be utilized in malware detection.

\vspace*{0.15in}
\noindent {\bf Road Map.} The remainder of this work is organized as follows. In  Section~2 we present and analyze our proposed model for malware detection based on group dependency graphs and describe the corresponding graph construction procedure. In  Section~3 we present a similarity metric that combines qualitative, quantitative, and relational characteristics. In  Section~4 we analyze our data set, describe our experimental design, evaluate our proposed model's implementation against real malware samples, and compare our results against the ones achieved by other models. Finally, in  Section~5 we conclude our paper and discuss possible future extensions.

\section{Model Design}

In this section, we leverage the so far unexploited grouping of system-calls, invoked by a program $\pi$, into groups of similar functionality and construct a graph that its vertex set consists of super-nodes containing the system-calls belonging to the same group, while its edge set contains the interconnection between the system-calls of these groups.

\subsection{The System-call Dependency Graph $G$}

It is well known that the actions performed by a program, depicting its behavior, rely on system-calls. Tracing the system-calls performed during the execution of a malware program $\pi$, we can represent its behavior interpreting this information with a graph, so called {\it System-call Dependency Graph} (or, ScD for short); throughout the paper, we shall denote a ScD graph by $D[\pi]$ and the system-calls invoked by $\pi$ by $S_i$, $1 \leq i \leq n$.

The vertex set of a ScD graph $D[\pi]$ is consisted by all the system-calls that take place during the execution of a program, i.e., $S_1$, $S_2$, $\ldots$, $S_n$, while its edge set contains the pairs of system-calls that exchanged arguments during the execution. Thus, an edge of ScD graph $D[\pi]$ is a tuple of type $(S_{i}$:$k$, $S_{j}$:$\ell)$ indicating that the system-call $S_{i}$ invokes $S_{j}$ and the $k^{th}$ output argument of $S_{i}$ is passed as the $\ell^{th}$ input argument of $S_{j}$.

\begin{table}[!h]
    \begin{minipage}[c]{2.7in}
      \centering
       \begin{tabular}{|c|l|c|c|}
		\hline
		\textbf{ID} & \textbf{\ System-call Name} &\textbf{In} & \textbf{Out} \\ \hline \hline
		0 & \ NtOpenSection \  & 2 & 1 \\ \hline
		1 & \ ACCESS\_MASK \  & 0 & 1 \\ \hline
		2 & \ POBJECT\_ATTRIBUTES \  & 0 & 1 \\ \hline
		3 & \ NtQueryAttributesFile \  & 1 & 1 \\ \hline
		4 & \ NtRaiseHardError \  & 5 & 0 \\ \hline
		5 & \ NTSTATUS \  & 0 & 1 \\ \hline
		6 & \ ULONG \  & 0 & 1 \\ \hline
		7 & \ PULONG\_PTR \  & 0 & 1 \\ \hline
		8 & \ HARDERROR\_RESPONSE\_OPTION \  & 0 & 1 \\ \hline
	\end{tabular}
	\vspace{0.1in}
    \end{minipage}
    \begin{minipage}[c]{2.7in}
      \centering
    	\vspace{-0.07 in}
        \begin{tabular}{|c|c|}
		\hline
		\textbf{Trace} & \textbf{Edge}\\ \hline \hline
		 \ 1:1,0:1 \  &  $S_{1} \longrightarrow S_{0} $\\ \hline
		 \ 2:1,0:1 \  &  $S_{2} \longrightarrow S_{0}$ \\ \hline
		 \ 2:1,3:1 \  & $S_{2} \longrightarrow S_{3} $\\ \hline
		 \ 2:1,3:1 \  & $S_{2} \longrightarrow S_{3} $\\ \hline
		 \ 2:1,3:1 \  & $S_{2} \longrightarrow S_{3} $\\ \hline
		 \ 2:1,3:1 \  & $S_{2} \longrightarrow S_{3} $\\ \hline
		 \ 2:1,3:1 \  & $S_{2} \longrightarrow S_{3} $\\ \hline
		 \ 2:1,3:1 \  & $S_{2} \longrightarrow S_{3} $\\ \hline
		 \ 2:1,3:1 \  & $S_{2} \longrightarrow S_{3} $\\ \hline
		 \ 2:1,3:1 \  & $S_{2} \longrightarrow S_{3} $\\ \hline
		 \ 2:1,3:1 \  & $S_{2} \longrightarrow S_{3} $\\ \hline
		 \ 2:1,3:1 \  & $S_{2} \longrightarrow S_{3} $\\ \hline
		 \ 5:1,4:1 \  & \  $S_{5} \longrightarrow S_{4} $ \ \\ \hline	
		 \ 6:1,4:2 \  & \  $S_{6} \longrightarrow S_{4} $ \  \\ \hline
		 \ 6:1,4:3 \  & \  $S_{6} \longrightarrow S_{4} $ \ \\ \hline
		 \ 7:1,4:4 \  & \  $S_{7} \longrightarrow S_{4} $ \ \\ \hline
		 \ 8:1,4:5 \  &  \ $S_{8} \longrightarrow S_{4}$ \ \\ \hline
		\end{tabular}
		\vspace{0.19in}
    \end{minipage}

	\centering  \hspace{0.7 in} {\bf (a)} \hspace{2.3 in} {\bf (b)} \vspace{0.19in}
	\vspace*{-0.1 in}
	\caption{(a) System-calls appeared during the execution of a program $\pi$ from a malware family Hupigon, (b) System-call dependencies \cite{Data-Set}.}
\label{Tab1}
\end{table}

\begin{figure}[t!]
    \hrule\medskip\smallskip
    \centering
    \includegraphics[scale=0.50]{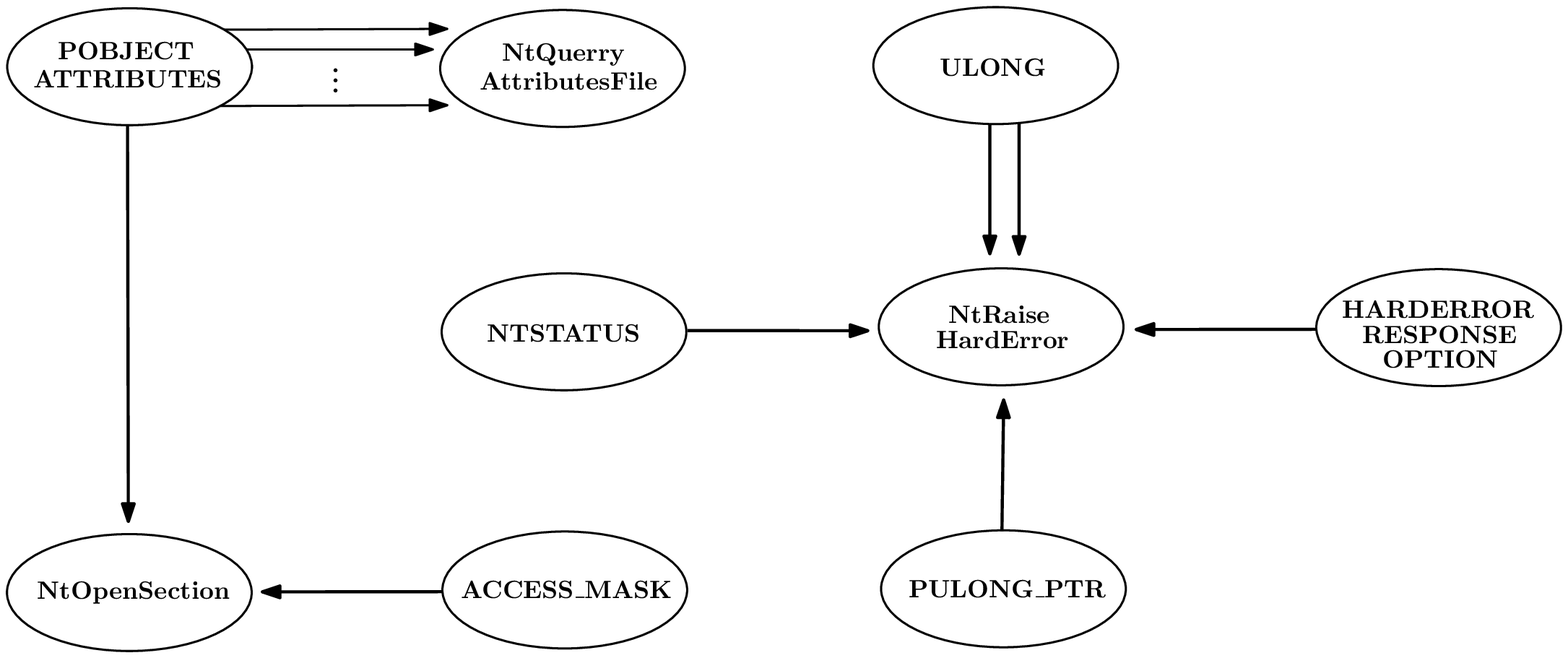}
    \centering
    \smallskip\medskip\hrule\medskip
    \caption{\small{A system-call dependency graph $D[\pi]$ of a program $\pi$.}}
\label{fig:fig1}
\end{figure}

Next, we illustrate a simple example that includes the system-call traces obtained through dynamic taint analysis \cite{BaReSo} during the execution of a sample from malware family Hupigon, downloaded from Domagoj Babic's personal webpage~\cite{Data-Set}, and we explain how the ScD graph is constructed after the whole process. Observing the data from the Table~\ref{Tab1}, we can see the construction of the ScD graph $D[\pi]$ that is a directed acyclic graph (dag); see, Figure~\ref{fig:fig1}. It is easy to see that the vertex set of this graph is consisted from the system-calls appeared during the execution of the sample and its edge set is consisted by their in between data-flow dependencies; see, Tables~\ref{Tab1}(a) and \ref{Tab1}(b).

Finally, we recall a well known fact that is the suspicious sample needs to be executed in a contained environment (i.e., a virtual machine), where during its execution time, taint analysis is performed in order to capture system-call traces.

\subsection{The Group Dependency Graph $D^*[\pi]$}

The key idea of our detection model is based on the, so far unexploited, information that system-calls of program $\pi$ of similar functionality can be classified into the same group, as we firstly presented it in \cite{ChNiPo}. For a proper system-call grouping we utilized the grouping provided by NtTrace \cite{NtTrace}, a system-call monitoring tool for MS Windows, complying with Micorsoft's documentation, where each system-call has a detailed description indicating the group it belongs to; we denote by $\mathcal{C}^*$ the set of system-call groups for a given operating system and by $\mathcal{C}_1$, $\mathcal{C}_2$, $\ldots$, $\mathcal{C}_{n^*}$ the groups of $\mathcal{C}^*$.

Thus, if a system-call dependency graph $D[\pi]$ of a given program $\pi$ is composed by $n$ system-calls $S_1$, $S_2$, $\ldots$, $S_n$, then each system-call $S_i$, $1 \leq i \leq n$, belongs to exactly one group $\mathcal{C}_j$, $1 \leq j \leq n^*$.

Having the grouping $\mathcal{C}^*$ and a system-call dependency graph $D[\pi]$, we next construct the key component of our model that is the {\it Group Dependency Graph} (or, GrD for short). The GrD graph, which we denote by $G^*[\pi]$, is a directed weighted graph on $n^*$ nodes $u_1$, $u_2$, $\ldots$, $u_{n^*}$; it is constructed as follows:

\begin{itemize}
\item[(i)] we first define a bijective function $f: V(G^*[\pi]) \longrightarrow \mathcal{C}_i$ from the node set $V(G^*[\pi])$ $=$ $\{u_1, u_2, \ldots, u_{n^*}\}$ to the set of groups $\mathcal{C}_i$ $=$ $\{\mathcal{C}_1, \mathcal{C}_2, \ldots, \mathcal{C}_{n^*}\}$;

\item[(ii)] for every pair of nodes $\{u_i, u_j\} \in V(G^*[\pi])$, we add the directed edge~$(u_i, u_j)$ in $E(G^*[\pi])$ if $(S_p,S_q)$ is an edge in $E(G[\pi])$ and, $S_p \in \mathcal{C}_i$ and $S_q \in \mathcal{C}_j$, $1 \leq i, j \leq n^*$;

\item[(iii)] for each directed edge $(u_i, u_j) \in E(G^*[\pi])$, we assign the weight $w$ if there are $w$ invocations from a system-call in group $f(u_i)= \mathcal{C}_i$ to a system-call in group $f(u_j)= \mathcal{C}_j$, $1 \leq i, j \leq n^*$.
\end{itemize}

\begin{figure}[t!]
    \hrule\medskip\smallskip
    \centering
    \includegraphics[scale=0.50]{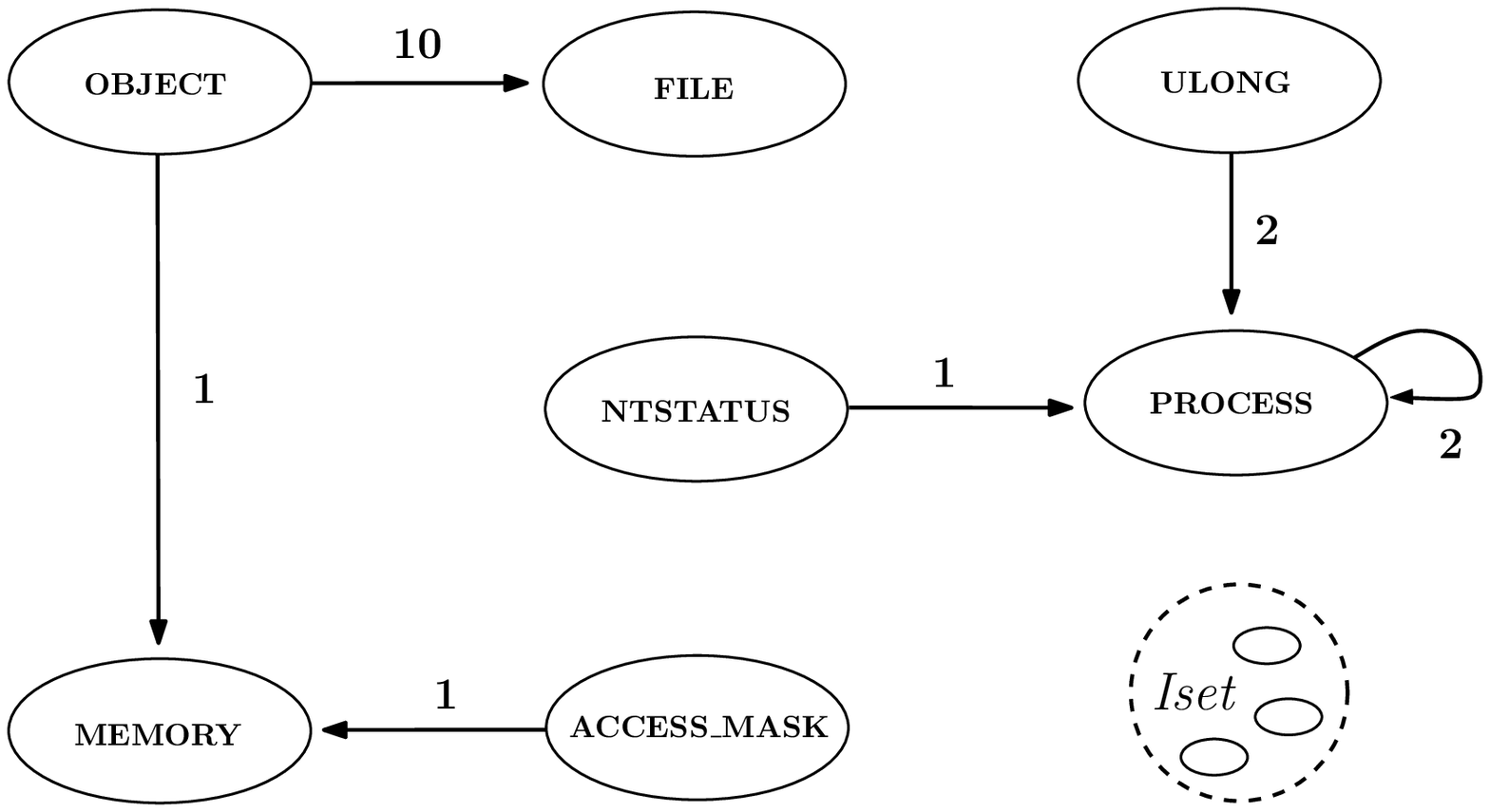}
    \centering
    \smallskip\medskip\hrule\medskip
    \caption{\small{The group dependency graph $D^*[\pi]$ of the graph of Figure~\ref{fig:fig1}.}}
\label{fig:fig2}
\end{figure}

\noindent Figure~\ref{fig:fig2} depicts the GrD graph~$D^*[\pi]$ of the ScD graph~$D[\pi]$ of Figure~\ref{fig:fig1}; the set $Iset$ contains all the isolated nodes of $D^*[\pi]$.

In Table~\ref{Tab2}, we present the groups of system-calls $\mathcal{C}_1$, $\mathcal{C}_2$, $\ldots$, $\mathcal{C}_{n^*}$ and the number of system-calls inside each group. We point out that the number of non-isolated nodes of graph $D^*[\pi]$ equals the number of groups formed by the system-call of graph $D[\pi]$; note that, the total number of nodes of $D^*[\pi]$ is always $n^*$. For example, the $9$ system-calls of ScD graph belong to $7$ groups (see, Table~\ref{Tab3}), the ScD graph $D[\pi]$ of Figure~\ref{fig:fig1} contains $9$ nodes, while its corresponding GrD graph $D^*[\pi]$ contains $7$ non-isolated nodes and thus $23$ isolated nodes in $Iset$; see, Figure~\ref{fig:fig2}.

It is extremely important to point out that while the ScD graph $D[\pi]$ is by definition an acyclic directed graph, the produced GrD graph $D^*[\pi]$ is not, in general, acyclic. As easily one can see that by grouping nodes in $D[\pi]$ it is very likely to create directed circles and/or self-loops; an indicative example appears in graph $D^*[\pi]$ of Figure~\ref{fig:fig2}.

\begin{table}[t!]
\centering
       \begin{tabular}{|l|c|l|c|}
\hline
\textbf{\ Group Name} & \textbf{Size} & \textbf{\ Group Name} & \textbf{Size} \\ \hline \hline
\ ACCESS\_MASK \phantom{XX}	&	1	&	\ PHANDLE		&	1	\\	\hline
\ Atom	&	5	& \ PLARGE\_INTEGER		&	1	\\	\hline
\ BOOLEAN	&	1	& \ Process		&	49	\\	\hline
\ Debug	&	17	& \ PULARGE\_INTEGER	\phantom{XX}	&	1	\\	\hline
\ Device	&	31	& \ PULONG		&	1	\\	\hline
\ Environment	&	12	& \ PUNICODE\_STRING		&	1	\\	\hline
\ File	&	44	& \ PVOID\_SIZEAFTER		&	1	\\	\hline
\ HANDLE	&	1	& \ PWSTR		&	1	\\	\hline
\ Job	&	9	& \ Registry		&	40	\\	\hline
\ LONG	&	1	& \ Security		&	36	\\	\hline
\ LPC	&	47	& \ Synchronization		&	38	\\	\hline
\ Memory	&	25	& \ Time		&	5	\\	\hline
\ NTSTATUS	&	1	& \ Transaction		&	49	\\	\hline
\ Object	&	19	& \ ULONG		&	1	\\	\hline
\ Other	&	36	& \ WOW64		&	19	\\	\hline
\end{tabular}  
\medskip\medskip
\caption{The 30 system-call groups.}
\label{Tab2}
\vspace*{-0.2 in}
\end{table}

\begin{table}[b!]	
\centering
\begin{tabular}{|c|l|c|}
		\hline
		\textbf{ID} & \textbf{\ System-call} & \textbf{Group} \\ \hline \hline
		0 & \ NtOpenSection \  & Memory \\ \hline
		1 & \ ACCESS\_MASK \  & \ ACCESS\_MASK \ \\ \hline
		2 & \ POBJECT\_ATTRIBUTES \  & Object \\ \hline
		3 &\ NtQueryAttributesFile \  & File\\ \hline
		4 &\ NtRaiseHardError \  & Process\\ \hline
		5 & \ NTSTATUS \  & NTSTATUS \\ \hline
		6 & \ ULONG \  & ULONG \\ \hline
		7 & \ PULONG\_PTR \  & Process\\ \hline
        8 & \ HARDERROR\_RESPONSE\_OPTION  \  & Process \\ \hline
		\end{tabular}
        \medskip\medskip
	\caption{The 9 system-calls of Figure~\ref{fig:fig1} and their corresponding groups.}
\label{Tab3}
\end{table}

\subsection{Family Identity Matrix}
\label{sec:53}

In this section we will describe the construction of an informative adjacency matrix that will act as a unique identity for each malware family. Our approach is based on the intuition that malware samples belonging to an individual malware family tend to share common characteristics. This is a quite valuable information, that we leveraged in order to develop a technique that will utilize these characteristics in order to decide if an unknown sample is malware or not.

Defining the term {\it characteristic} when working on GrD graph $D^*[\pi]$, we could claim that a characteristic is an edge between two system-call groups, since in order for an individual task to be performed, system-calls of specific functionality need to be utilized and of course in different malware variants they can be substituted by equivalent ones. Thus, we decided to focus on edges that exist in most of the members' GrD graphs $D^*[\pi]$, constituting hence a qualitative characteristic of their family.

So, easily one can understand that, if in a malware family, a specific edge, appears in the majority of the members, then this edge exposes a greater significance, in contrast with another one that exists in the minority of the members of this family. Hence, in order to represent the significance of an edge we take into account the percentage of the members in a family in which this edge has a non-zero value. To this point we ought to underline that, since the values in the cells of adjacency matrix refer to the weight $w$ of the corresponding edge, in order to claim about the significance of an edge as a qualitative characteristic of a family, we are interested only on the non-zero weights.

\begin{figure}[t!]
  \hrule\vspace*{0.3cm}
  \centering
    \includegraphics[scale=0.50]{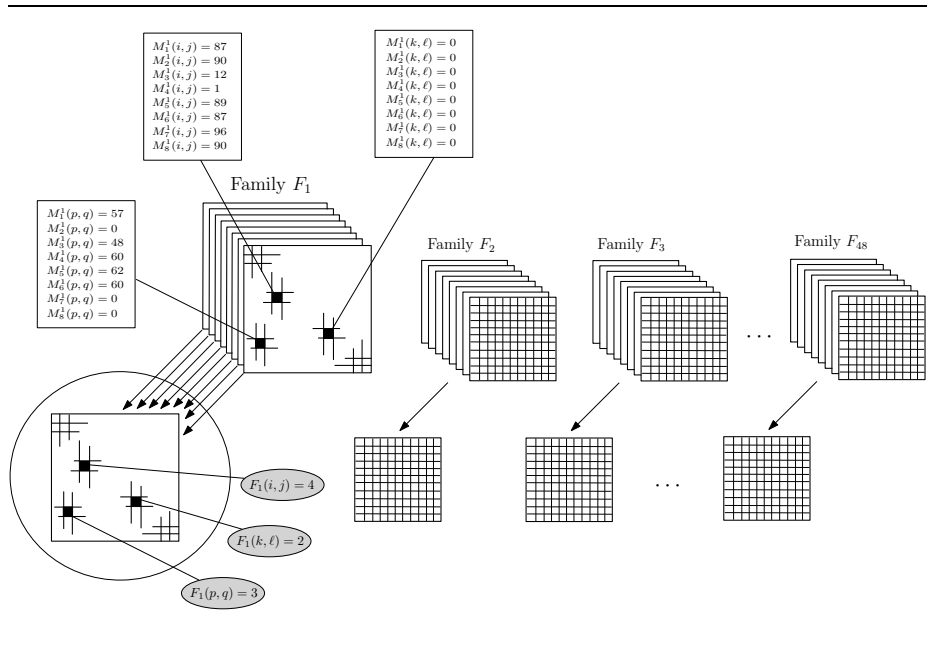}
    \centering
    \medskip\hrule\medskip
    \caption{\small{The structure of the ID-matrix.}}
    \medskip
\label{FamilyID}
\end{figure}

Thus, having collected this valuable information we proceed by filtering it as to decide the significant edges that will indicate the characteristics of each family. Hence, having computed the percentage of appearance of each edge we can assign weights to each cell (i.e., edge) on this matrix constructing hence the {\it ID-matrix} of the family as shown in Figure~\ref{FamilyID}.

Finally, in order to assign weights we partition the values (ranging from 0 to 100) to three categories. However, before we assign the significance tags we ought to define the value ranges. So, we first define a threshold about 95$\%$ and the tags are arranged based on this threshold. Thus, we mark each cell with a tag either {\tt Red}, {\tt Gray}, or {\tt White}, with value {\tt 4}, {\tt 3}, or {\tt 2}, respectively.

In our model, the {\tt Red} tags cover cells containing values in the range $[0.95-1]$, the {\tt Gray} tags cover cells containing values in the range $(0.05-0.95)$, while the {\tt White} tags cover the ones containing values in the range $[0-0.05]$.

\section{Model Metrics}
In this section, we present a similarity metric that combines qualitative, quantitative, and relational characteristics that are spread among the members of known malware families to achieve a distinction between a malware and a benign software.

\subsection{GrD-Similarity Metrics}
\label{sec:GrDSM}

\noindent We next present the proposed methodology for malware detection. Specifically, we present the computation of the similarity between a test sample $T$ and every malware family $F_{k}$ of the set $\mathcal{F}$, $1 \leq k\leq |\mathcal{F}|$. Before we proceed with the presentation of the similarity metrics let us give some definitions and notations used throughout the paper.

\begin{definition}
Let $A$, $B$ two $n \times n$  matrices with elements $A(\cdot)$, $B(\cdot) \in \mathrm{R}$, and let $p, q \in \mathrm{Z}$. Then, we define
\end{definition}

\begin{equation}
|A(i,j) \cap_{p\rightarrow q} B(i,j)|=
\begin{cases}
    1, & \mbox{if } A(i,j)=p \mbox{ and } B(i,j)=q, \  1 \leq i,j \leq n \\
    0, & \mbox{otherwise}
\end{cases}
\end{equation}

\noindent and

\begin{equation}
|A \cap_{p\rightarrow q} B\vert = \sum\limits_{i=1}^{n}\sum\limits_{j=1}^{n}\vert A(i,j) \cap_{p\rightarrow q} B(i,j)|.
\end{equation}

\vspace*{0.1in}
\noindent By definition, $|A\cap_{p\rightarrow p} A|$ is the number of the elements of matrix $A$ with value $p$; hereafter, this number is referred to as $|A=p|$.

Next, we provide a set of GrD graph similarity metrics along with a description of their qualitative, quantitative, and relational characteristics. The following adjacency matrices represent the GrD graph $D^*[\pi]$

\vspace*{0.2in}
\noindent {\bf A. Family-Test Similarity}
\vspace*{0.05in}

\noindent The Family-Test similarity metric focuses on the computation of the similarity between the test sample and any malware family using the family's ID-matrix. We distinguish two similarity approaches, namely

\begin{itemize}
\item[$\circ$] Family.to.Test cover, and
\item[$\circ$] Test.to.Family cover.
\end{itemize}

\noindent The main purpose of the Family.to.Test cover similarity metric is to compute the rate of satisfiability on the qualitative characteristics of any family of the set $\mathcal{F}$ by a test sample, while the main purpose of the Test.to.Family cover similarity metric is to compute the rate of satisfiability, in terms of edge existence, of a test sample's adjacency matrix by the qualitative characteristics of any family represented by its ID-matrix.

In order to compare the test sample's adjacency matrix $T$ with the ID-matrix $F_k$ of the $k$th family of the set $\mathcal{F}$, we first need to make a cast on test sample's adjacency matrix $T$. Recall that, the cells of the test sample's adjacency matrix have either zero or non-zero values. Thus, we cast any non-zero values existed in test sample's adjacency matrix $T$ into 1s resulting the casted matrix $T^{c}$.

\vspace*{0.1in}
\noindent {\bf Family.to.Test cover:} The main process of this similarity metric is to cover in some fashion some qualitative characteristics of family $F_k$ with the edge existence of the test sample. We achieve such a coverage by first computing the number of cells with a Red tag or, equivalently, a value 4 in family's ID-matrix $F_k$ that their corresponding cells in the test sample's casted matrix $T^{c}$ have value 1, i.e., $F_{k}(i,j)=4$ and $T^{c}(i,j)=1$, and then dividing this number by the total number of cells in family's ID-matrix that have a Red tag, i.e., $F_{k}(i,j)=4$.

We call the above proposed cover {\it Four.to.One cover} and denote it, in a form of function, as $Four.to.One(F_{k},T^{c})$. Thus, the formula that gives the Family.to.Test cover similarity metric is the following:

\begin{equation}
    Four.to.One(F_{k},T^{c})=\dfrac{\vert F_{k} \cap_{4\rightarrow 1} T^{c} \vert}{ \vert F_{k}=4 \vert}
\end{equation}

\noindent where, $F_{k}$ is the ID-matrix of the $k^{th}$ family of a set $\mathcal{F}$ of families and $T^{c}$ is the test sample's casted adjacency matrix.

\vspace*{0.1in}
\noindent {\bf Test.to.Family cover:} This metric computes the satisfiability of the edge existence in the test sample, represented by the topology of 1s in the casted matrix $T^{c}$, by the qualitative characteristics of a family $F_k$, represented by the topology of Red tags in its ID-matrix. In a similar way, we achieve a Test.to.Family coverage by first computing the number of cells in the test sample's casted matrix $T^{c}$ having value 1 that their corresponding cells in family's ID-matrix $F_k$ have a Red tag or, equivalently, a value 4, i.e., $T^{c}(i,j)=1$ and $F_{k}(i,j)=4$, , and then dividing this number by the total number of cells in the test sample's casted matrix $T^{c}$ that having value 1, i.e., $T^{c}(i,j)=1$.

As above, we also call the proposed cover {\it One.to.Four cover} and denote it, in a form of function, as $One.to.Four(T^{c},F_{k})$. Thus, the formula that gives the Test.to.Family cover similarity metric is the following:

\begin{equation}
One.to.Four(T^{c},F_{k})=\dfrac{\vert T^{c} \cap_{1\rightarrow 4} F_{k} \vert}{\vert T^{c}=1 \vert}
\end{equation}

\noindent where, again $F_{k}$ is the ID-matrix of the $k^{th}$ family of a set $\mathcal{F}$ of families and $T^{c}$ is the test sample's casted adjacency matrix.

\vspace*{0.2in}
\noindent {\bf B. Jaccard Similarity}
\vspace*{0.05in}

\noindent One more similarity metric we utilize to empower our formula for malware detection is the Jaccard index \cite{jaccard_tanimoto}. The reason we choose to utilize the Jaccard similarity is the fact that it is mostly applied on binary vectors and thus it seems to efficiently work for the comparison between two graph-objects in terms of edge existence. More precisely, this metric measures the similarity of relational characteristics, in terms of edge existence, between the test sample's casted matrix $T^{c}$ and a member's casted matrix $M^{c}$ of a malware family $F_{k}$.

We first utilize the Jaccard index to compute the maximum value produced by the most similar member of the family to the test sample, and then we compute the mean similarity between the test sample and all the members of a malware family .

The computation of the Jaccard similarity metric is achieved by first computing the number of cells that have value 1 in the test sample's casted matrix $T^{c}$ and their corresponding cells in the member's casted matrix $M^{c}$ have also value 1, i.e., $T^{c}(i,j)=1$ and $M^{c}(i,j)=1$, and then dividing this number by the number of the cells that either in $T^{c}$ or in $M^{c}$ have value 1, i.e., $T^{c}(i,j)=1$ or $M^{c}(i,j)=1$. Hence the computation of Jaccard similarity can be computed as follows:

\begin{equation}
J(T^{c},M^{c})=\dfrac{\vert T^{c} \cap_{1\rightarrow 1} M^{c} \vert }{ \vert T^{c} \cap_{1\rightarrow 1}M^{c}\vert + \vert T^{c} \cap_{1\rightarrow 0}M^{c} \vert + \vert T^{c} \cap_{0\rightarrow 1}M^{c} \vert} ,
\end{equation}

\vspace*{0.06in}
\noindent where $M^{c}$ is the member's casted adjacency matrix and $T^{c}$ is the test sample's casted adjacency matrix.

Let $F_{k}$ be a malware family containing $m_k$ members and let $M^{c}_{1}, M^{c}_{2}, \ldots, M^{c}_{m_k}$ be the casted matrices of the members of $F_{k}$. The maximum Jaccard similarity, produced by the most similar member of $F_{k}$ to the test sample, is defined as follows:

\begin{equation}
J_{\text{max}}(T^{c},F_{k})=\max\limits_{1\leq \ell \leq m_{k}}[J(T^{c},M^{c}_{\ell})],
\end{equation}

\vspace*{0.06in}
\noindent while the mean Jaccard similarity, produced by the mean of all the Jaccard similarity values between the test sample and the members of $F_{k}$, is defined by the following formula:

\begin{equation}
J_{\text{mean}}(T^{c},F_{k})=\dfrac{\sum\limits_{\ell=1}^{m_{k}}J(T^{c},M^{c}_{\ell})}{m_{k}},
\end{equation}

\noindent where $F_{k}$ is the $k^{th}$ malware family of a set $\mathcal{F}$ and $T^{c}$ is the test sample's casted adjacency matrix.

\vspace*{0.2in}
\noindent {\bf C. Bray-Curtis Similarity}
\vspace*{0.05in}

\noindent The last similarity metric we utilize to empower our formula for malware detection is the Bray-Curtis dissimilarity \cite{Bray_Curtis}. The reason we select the Bray-Curtis dissimilarity is the fact that it is mostly applied for the computation of diversity between two object represented by vectors of continuous values. More precisely, this metric measures the similarity of quantitative characteristic, in terms of edge weights, between the test samples' adjacency matrix $T$ and a member's adjacency matrix $M$. However, since the return value of the Bray-Curtis metric refers to the distance between any two objects, it is in the range $[0,1]$ with maximum value the 0. Thus, in order to reverse this property we perform a subtraction from 1.

As with the Jaccard similarity, we also utilize the Bray-Curtis similarity to compute first the maximum value produced by the most similar member of the family to the test sample, and then we compute the mean similarity between the test sample and all the members of a malware family.

The computation of Bray-Curtis similarity is achieved by computing the sum of subtractions of the corresponding values of cells $T(i,j)$ and $M(i,j)$ and dividing this number by the sum of their additions. Hence, the Bray-Curtis similarity is given by the following formula:

\begin{equation}
BC(T,M) =1-\dfrac{\sum\limits_{i=1}^{n}\sum\limits_{j=1}^{n} (T(i,j) - M(i,j))}
{\sum\limits_{i=1}^{n}\sum\limits_{j=1}^{n} (T(i,j) + M(i,j))} ,
\end{equation}

\noindent where $T$ is the $n \times n$ adjacency matrix of test sample and $M$ is the $n \times n$ adjacency matrix of the member of family $F_{k}$ under consideration.

Next, we first define the maximum Bray-Curtis similarity, produced by the most similar member of $F_{k}$ to the test sample, as follows:

\begin{equation}
BC_{\text{max}}(T,F_{k})=\max\limits_{1\leq \ell \leq m_{k}}[BC(T,M_{\ell})] ,
\end{equation}

\noindent and then we define the mean Bray-Curtis similarity, produced by the mean of all the Bray-Curtis similarity values between the test sample and the members of $F_{k}$, as follows:

\begin{equation}
BC_{\text{mean}}(T,F_{k})=\dfrac{\sum\limits_{\ell=1}^{m_{k}}BC(T,M_{\ell})}{m_{k}} ,
\end{equation}

\noindent where $F_{k}$ is the $k^{th}$ malware family of the set $\mathcal{F}$, $m_{k}$ is the number of members of the family $F_{k}$, $M_{\ell}$ is the $\ell^{th}$ member's  adjacency matrix and $T$ is the test sample's adjacency matrix.

\vspace*{0.2in}
\noindent {\bf D. Tanimoto Similarity}
\vspace*{0.05in}

\noindent Finally we use the Tanimoto similarity \cite{jaccard_tanimoto} for filtering purposes in our proposed malware detection model. The Tanimoto similarity is a mechanism for computing the Jaccard coefficient when the set under comparison are represented as bit vectors.

As with the previously described similarity metrics, we also utilize the Tanimoto similarity to compute the maximum value produced by the most similar member of the family to the test sample.

The computation of Tanimoto similarity is achieved by computing the sum of subtractions of the corresponding values of cells $T(i,j)$ and $M(i,j)$ and dividing this number by the sum of their additions. Hence, the Tanimoto similarity is given by the following formula:

\begin{equation}
TN(A,B)=\dfrac{\sum\limits_{i=1}^{n} (T(i,j) \times M(i,j))}{\sum\limits_{i=1}^{n} (T(i,j))^2 + \sum\limits_{i=1}^{n} (T(i,j))^2 - \sum\limits_{i=1}^{n} (T(i,j) \times M(i,j))} ,
\end{equation}

\noindent where $T$ is the $n \times n$ adjacency matrix of test sample and $M$ is the $n \times n$ adjacency matrix of the member of family $F_{k}$ under consideration.

Next, we define the max Tanimoto similarity, produced by the most similar member of $F_{k}$ to the test sample, as follows:

\begin{equation}
TN_{\text{max}}(T,F_{k})=\max\limits_{1\leq \ell \leq m_{k}}[TN(T,M_{\ell})] ,
\end{equation}

\noindent where $F_{k}$ is the $k^{th}$ malware family of the set $\mathcal{F}$, $m_{k}$ is the number of members of the family $F_{k}$, $M_{\ell}$ is the $\ell^{th}$ member's  adjacency matrix and $T$ is the test sample's adjacency matrix.

\subsection{NP-similarity Metric}
\label{sec:NP}

Having presented several variants of Family-Test, Jaccard, and Bray-Curtis similarity metrics, let us now describe the NP-similarity metric which we have developed in order to detect whether an unknown test sample is a malicious or a benign one. More precisely, this metric globally measures the similarity between a test sample and a malware family combining, in a specific manner, the aforementioned similarity metrics taking into account the qualitative, the quantitative, and the relational characteristics of the objects under consideration.

The NP-similarity incorporates a combination of the similarity metrics referenced previously, where their contribution to its final result is affected by assigning different weights to each one of these similarity metrics. For our purpose, we choose four factors $a$, $b$, $c_{1}$, and $c_{2}$, and define three similarity-components namely $F_1$, $F_2$, and $F_3$.

The first similarity-components $F_1$ of our NP-similarity metric concerns the qualitative characteristics. Thus, we utilize the
Four.to.One cover similarity along with the One.to.Four similarity assigning greater weight factors in the Four.to.One. Indeed, we choose the factor $a=4$ for the Four.to.One similarity and the factor $b=2$ for the One.to.Four, while we choose greater weight factor for the case where both $Four.to.One()$ and $One.to.Four()$ take the maximum value 1; we express our choice by the following function:

\begin{equation*}
\phi= a \cdot Four.to.One(F_{k},T^{c}) + b \cdot One.to.Four(T^{c},F_{k})
\end{equation*}

\noindent Our choice is based on the intuition that, if the test sample is malicious, then it should be an expansion of a malware family inheriting and hence satisfying its qualitative characteristics. Moreover, the reason that we multiply the $Four.to.One()$ similarity by a greater factor is the fact that when this similarity metric is maximized it is indicating that the sample is a direct extension of the malware family.

Additionally, in the case where $Four.to.One()=1$ and $One.to.Four()=1$, the topology of 4s in Family ID-Matrix is identical to the topology of 1s in test samples casted matrix and thus we multiply the function $\phi$ by the factor $c_{1}=1.5$, otherwise we multiply it by the factor $c_{2}=1.2$. Hence, we define the first similarity-components $F_1$ of our NP-similarity metric as follows:

\begin{equation}
F_{1}=
\begin{cases}
    c_1 \cdot (a + b), & \mbox{if $Four.to.One(F_{k},T^{c})=One.to.Four(T^{c},F_{k})=1$} \\
    c_2 \cdot \phi,           & \mbox{otherwise}
\end{cases}
\end{equation}

\noindent where, $a=4$, $b=2$, $c_{1}=1.5$ and $c_{2}=1.2$, while $F_{k}$ is the $k^{th}$ malware family of a set $\mathcal{F}$ and $T^{c}$ is the test sample's casted adjacency matrix.

The second similarity-components $F_2$ of our NP-similarity metric measures the similarity of relational
characteristics between the test sample and a malware family as described by the Jaccard index.
We assign appropriate weights on the max and mean Jaccard similarities, i.e., $J_{\text{max}}()$ and $J_{\text{mean}}()$, by using the factors $a$ and $b$, as follows:

\begin{equation}
F_{2}=a \cdot J_{max}(T^{c},F_{k}) + b \cdot J_{mean}(T^{c},F_{k})
\end{equation}

\noindent where, $a=4$ and $b=2$.

We next proceed by defining the third similarity-components $F_3$ that measures the similarity of qualitative
characteristics assigning, as before, appropriate weights on the max and mean Bray-Curtis similarities as follows:

\begin{equation}
F_{3}=a \cdot BC_{max}(T,F_{k}) + b \cdot BC_{mean}(T,F_{k})
\end{equation}

\noindent where, $a=4$ and $b=2$.

We point out that in similarity-components $F_{2}$ and $F_{3}$ we assign a greater weight to the max Jaccard and max Bray-Curtis since, as we describe above, it is more likely for the test sample to be a direct mutation of a member of a malware family in the case where it is a malware.

We finally define our NP-similarity metric by combining the three similarity-components $F_1$, $F_2$, and $F_3$, as follows:

\begin{equation}
NP(F_{k},T)=\dfrac {F_{1} \cdot F_{2} \cdot F_{3}}{Q},
\end{equation}

\noindent where, $Q$ is a normalization factor equals the maximum value of the product $F_1 \cdot F_2 \cdot F_3$ so that $NP() \in [0,1]$.

\bigskip
\noindent \textbf{Intuition.} The whole process of the NP-similarity construction, by the aspect of weights assignment, is based on the intuition that during the polymorphism procedure, it is more probable for a new strain to be a direct mutation from a member of a malware family. Hence, so for first component, in the case of the Four.to.One similarity, as for the other two, in the cases of max Jaccard and max Bray-Curtis similarities respectively, we assign a greater weight on to them as to emphasize that probability.

\subsection{Malware Detection using NP Similarity}
\label{sec:MDGrD}

Next, we show how can utilize the NP-similarity for malware detection based on GrD graphs; recall that, for a malware program $\pi$ the GrD graph is denoted by $D^*[\pi]$. The methodology we follow is simple: given a test sample $T$, we compute the NP-similarity metric between $T$ and all the malware families $F_{1}, F_{2}, \ldots, F_{n}$ of a set $\mathcal{F}$, and then we accordingly compute the max Tanimoto similarity exhibited by a member of each family. Finally, for the families with the maximum Tanimoto similarity, we check the one with the corresponding maximum NP-similarity and if this maximum value, indicating the most similar family $F_{k}$ to $T$ according to NP-similarity, is above the specified threshold $\lambda$, i.e., $NP(F_{k},T)~\geq~\lambda$, we claim that the test sample belongs to $\mathcal{F}$ and thus it is a malware.

It is worth noting to mention that, as we will discuss later, it is experientially proven that the application of the NP-similarity archives a satisfying distinction between the GrD graphs $D^*[\pi]$ representing malware and those ones representing benign software.

\section{Evaluation}

In this section we first present our experimental design and discuss the reasons that we adopt the proposed evaluation setup. Then, we discuss how we divide our data set into train-set and test-set and how we tune our threshold parameters according to feedback produced by a series experiments. Finally, we present our detection results after the application of NP-similarity and compare our results with those of other models.

\subsection{Experimental Design}\label{sec:61}

In order to evaluate our proposed malware detection technique we use a dataset of $2631$ malware samples from a set $\mathcal{F}$ of $48$ malware families $F_1$, $F_2$, $\ldots$, $F_{48}$, each $F_k$ containing from $3$ to $317$ malware members, and also a set of $33$ benign samples.

Additionally, it is of major importance to mention that we do not perform any taint malware analysis on the samples due to the risk posed to the systems connected to the same network. Thus, we downloaded the initial System-call Dependency Graphs produced by taint analysis from the web-page of Domagoj~Babic~\cite{Data-Set} and transformed each sample's ScD graph $D[\pi]$  into GrD graph $D^*[\pi]$, based on the grouping of system-calls presented in Table~\ref{Tab2}. The set $\mathcal{F}$ of the $48$ malware families along with their sizes (i.e., number of members) are listed in Table~\ref{Tab6}.

\begin{table}[h!]
\begin{center}
\begin{tabular}{|l|c|l|c|}
\hline
\textbf{\ Family Name} & \textbf{Size} & \textbf{\ Family Name} & \textbf{Size} \\ \hline \hline

\ ABU,Banload & 16 & \ Hupigon,AWQ & 219 \\ \hline

\ Agent,Agent & 42 & \ IRCBot,Sdbot & 66 \\ \hline

\ Agent,Small & 15 & \ LdPinch,LdPinch & 16 \\ \hline

\ Allaple,RAHack & 201 & \ Lmir,LegMir & 23 \\ \hline

\ Ardamax,Ardamax & 25 & \ Mydoom,Mydoom & 15 \\ \hline

\ Bactera,VB & 28 & \ Nilage,Lineage & 24 \\ \hline

\ Banbra,Banker & 52 & \ OnLineGames,Delf & 11 \\ \hline

\ Bancos,Banker & 46 & \ OnLineGames,LegMir & 76 \\ \hline

\ Banker,Banker & 317 & \ OnLineGames,Mmorpg & 19 \\ \hline

\ Banker,Delf & 20 & \ OnLineGames,OnLineGames\phantom{xx} & 23 \\ \hline

\ Banload,Banker & 138 & \ Parite,Pate & 71 \\ \hline

\ BDH,Small & 5 & \ Plemood,Pupil & 32 \\ \hline

\ BGM,Delf & 17 & \ PolyCrypt,Swizzor & 43 \\ \hline

\ Bifrose,CEP & 35 & \ Prorat,AVW & 40 \\ \hline

\ Bobax,Bobic & 15 & \ Rbot,Sdbot & 302 \\ \hline

\ DKI,PoisonIvy & 15 & \ SdBot,SdBot & 75 \\ \hline

\ DNSChanger,DNSChanger \phantom{xx} & 22 & \ Small,Downloader & 29 \\ \hline

\ Downloader,Agent & 13 & \ Stration,Warezov & 19 \\ \hline

\ Downloader,Delf & 22 & \ Swizzor,Obfuscated & 27 \\ \hline

\ Downloader,VB & 17 & \ Viking,HLLP & 32 \\ \hline

\ Gaobot,Agobot & 20 & \ Virut,Virut & 115 \\ \hline

\ Gobot,Gbot & 58 & \ VS,INService & 17 \\ \hline

\ Horst,CMQ & 48 & \ Zhelatin,ASH & 53 \\ \hline

\ Hupigon,ARR & 33 & \ Zlob,Puper & 64 \\ \hline

\end{tabular}
\medskip\medskip
\caption{The set $\mathcal{F}$ of the $48$ malware families $F_1, F_2, \ldots, F_{48}$, along with their sizes, i.e., number of members, downloaded from~\cite{Data-Set}.}
\label{Tab6}
\end{center}
\vspace*{-0.2 in}
\end{table}

For evaluation purposes of our model, we perform $5$-fold cross validation utilizing the dataset we described above. Additionally, we set the detection threshold $\lambda=0.56$ (see, Section~\ref{sec:MDGrD}), after performing a number of experiments focusing on maximizing the ratio of true-positives by the false-positives.

\subsection{Detection Results}

Next, we present our results after performing a set of 5-fold cross validation experiments partitioning the data set described above into 5 buckets  using in each experiment one bucket as test-set and the other four as train-set. In Table~\ref{Tab7} we cite our results concerning the detection rates and the corresponding false positives for various values of threshold $\lambda$ as we described it previously. To this point we ought to notice that due to the 5-fold cross validation process the percentage values below are averaged over the five buckets.

\begin{table}[h!]
\begin{center}
\begin{tabular}{|c||c||c|}
\hline
\textbf{\ \ Threshold $\lambda$ \ \ }& \textbf{\ \ Detection Rate \ \ } & \textbf{\ \ False Positives \ \ }\\ \hline
$\lambda$ = 0.35  & 98.14 \% & 68.57 \% \\ \hline
$\lambda$ = 0.42  & 96.70 \% & 56.00 \% \\ \hline
$\lambda$ = 0.51  & 94.06 \% & 29.00 \% \\ \hline \hline
$\lambda$ = 0.56  & 91.32 \% & 13.70 \% \\ \hline \hline
$\lambda$ = 0.61  & 85.28 \% & 6.85 \% \\ \hline
$\lambda$ = 0.67  & 74.42 \% & 4.00 \% \\ \hline
$\lambda$ = 0.74  & 63.03 \%  & 2.28 \% \\ \hline
$\lambda$ = 0.81  & 39.64 \% & 0.00 \% \\ \hline
\end{tabular}
\medskip\medskip
\caption{Malware detection results.}\label{Tab7}
\end{center}
\vspace*{-0.2 in}
\end{table}

\begin{figure}[t!]
  \hrule\vspace*{0.3cm}
  \centering
    \includegraphics[scale=0.55]{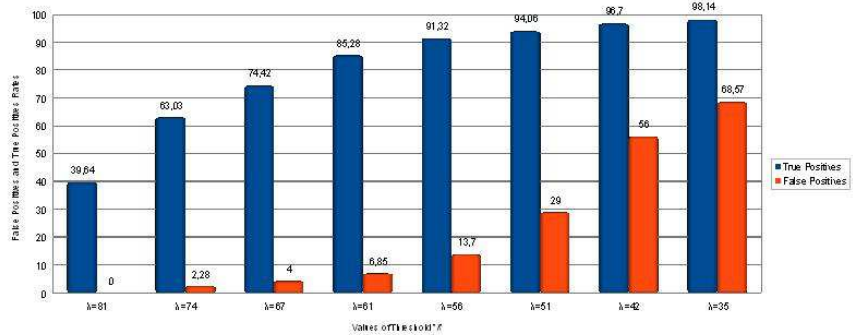}
    \centering
    \medskip\hrule\medskip
    \caption{\small{Detection results for multiple values of $\lambda$ depicting true and false positives variations.}}
\label{np_chart}
\end{figure}

In Figure \ref{np_chart} we provide an observation of the detection rates (True Positives) and False Positives of our proposed model, while changing the value of threshold $\lambda$. As mentioned previously, due to the $k$-fold cross validation process all the percentage values shown in the figures are averaged over the $k$-folds, (i.e., 5 folds for our proposed model).

In Table~\ref{Tab8}, we illustrate a comparison of our detection rates (i.e., true-positives) and the fail detections (i.e., false-positives) against those presented in other research works. We compare our results to graph-based and other techniques both using different data-sets. More precisely, the first column refers to the result's host, the second one refers to the utilized technique, while the third and fourth columns refer to the detection and false-positive rates, respectively.

\begin{table}[b!]
\begin{center}
\begin{tabular}{|r||l|c|c|}
\hline

\textbf{In:}\phantom{x} & \textbf{\ Technique} & \textbf{True Positives} & \textbf{False Positives} \\ \hline\hline

\cite{AlLaVeWa}\phantom{x} & \ SVM classifier (API-sequences) & 89.74 \% & 09.74 \% \\ \hline

\cite{YeDiTaDo}\phantom{x} & \ OOA rules (API-sequences) & 97.19 \% & 00.12 \% \\ \hline

\cite{ChJhSeSoBr}\phantom{x} & \ Templates (CFG) & 97.50 \% & 00.00 \% \\ \hline\hline

\cite{KoCoKrKiZhWa}\phantom{x} & \ Sequence Matching (ScD) & 64.00 \% & 00.00 \% \\ \hline

\cite{LuTa}\phantom{x} & \ Graph-Grading(ScD) & 80.09 \% & 11.00 \% \\ \hline

\cite{FRJhChSaYa}\phantom{x} & \ Graph Mining (ScD) & 92.40 \% & 6.1 \% \\ \hline

\cite{BaReSo}\phantom{x} & \ Tree Automata Inference (ScD) \phantom{x} & 80.00 \% & 05.00 \% \\ \hline

{this paper}\phantom{x} & \ NP-Similarity (GrD) & 91.32 \% & 13.70 \% \\ \hline

\end{tabular}
\medskip\medskip
\caption{Malware detection results comparison. Note that this paper uses the same dataset as \cite{BaReSo} and \cite{FRJhChSaYa}.}\label{Tab8}
\end{center}
\vspace*{-0.2 in}
\end{table}

Alazab {\it et~al.}~\cite{AlLaVeWa} developed a fully automated system that disassembles and extracts API-call features from executables and then, using $n$-gram statistical analysis, is able to distinguish malicious from benign executables. The mean detection rate exhibited was 89.74\% with 9.72\% false-positives when used a Support Vector Machine (SVM) classifier by applying $n$-grams.


Ye {\it et al.}~\cite{YeDiTaDo} described an integrated system for malware detection based on API-sequences. This is also a different model from ours since the detection process is based on matching the API-sequences on OOA rules (i.e., Objective-Oriented Association) in order to decide the maliciousness or not of a test program.

An important work of Christodorescu {\it et al.}, presented in~\cite{ChJhSeSoBr}, proposes a malware detection algorithm, called $A_{\text{MD}}$, based on instruction semantics. More precisely, templates of control flow graphs are built in order to demand their satisfiability when a program is malicious. Although their detection model exhibits better results than the ones produced by our model, since it exhibits 0 false-positives, it is a model based on static analysis and hence it would not be fair to compare two methods that operate on different objects.

Kolbitch {\it et al.}~\cite{KoCoKrKiZhWa} proposed an effective and efficient approach for malware detection, based on behavioral graph matching by detecting string matches in system-call sequences, that is able to substitute the traditional anti-virus system at the end hosts. The main drawback of this approach is the fact that although no false-positives where exhibited, their detection rates are too low compared with other approaches.

Luh and Tavolato~\cite{LuTa} present one more detection algorithm based on behavioral graphs that distinguishes malicious from benign programs by grading the sample based on reports generated from monitoring tools. While the produced false-positives are very close to ours, the corresponding detection ratio is even lower.

Fredrikson {\it et al.}~\cite{FRJhChSaYa} proposed an automatic technique for extracting optimally discriminative specifications based on graph mining and concept analysis that, when used by a behavior based malware detector, it can efficiently distinguish malicious from benign programs. The proposed technique can yield an 86.5\% detection rate with 0 false-positives. Since we compare only the maximum detection rates exhibitied by each technique, in Table \ref{Tab8} we show the maximum detection rate 99.4\% which however exhibits higher false-positives (57.14\%). However, a more fair comparison would be the one depicted in Figures \ref{np_chart} and \ref{other_chart} where for specific values of $t$ and $\lambda$ (i.e. $t=0.96$ and $\lambda=0.56$) someone can observe that our model reaches the detection rates of the proposed model presented in \cite{FRJhChSaYa} with barely $0.03\%$ more false positives, proving the potentials of our model in a further improvement.

Finally, Babic {\it et al.}~\cite{BaReSo} achieved the malware detection by $k$-testable tree automata inference from system-call data flow dependence graphs. To this point we ought to underline that in this work the authors use the same data-set that we borrow from Domagoj~Babic's web-page~\cite{Data-Set}. Thus, this work provides a fair instance to compare our model's results. However, while Babic {\it et al.} perform 2-fold cross validation using the first half of data-set as train-set and the second one as test-set, we perform 5-fold cross validation. Comparing the results exhibited in \cite{BaReSo} with ours, easily we can claim that our proposed model is quite competitive to Babic's especially for specific values of $\lambda$ ($0.61$ and $0.67$ respectively).


\section{Concluding Remarks}

We have presented an elaborated graph-based algorithmic technique for efficient malware detection by exploiting main properties of system-call dependency graphs. We leveraged the partitioning of system-calls in order to construct the GrD graph $D^*[\pi]$ that depicts the interconnection of specific groups of system-calls. Then, we developed the NP-similarity metric that, operating on GrD graphs, combines a set of similarity metrics in order to distinguish whether an unknown test sample is malicious or not based on a predefined threshold.

We evaluated our model's detection ability and compared its potentials against other results from several models either graph-based or not. The evaluation was performed on a set of $2630$ malware samples from $48$ malware families and $33$ benign commodity programs. The detection process exhibited a 91.3\% rate with 13.7\% false positives making it competing against other detection models.

Finally, an interesting perspective is the extension of our model for malware indexing, i.e., to classify a test sample in a malware family, if it has been detected as malware; we leave such an extension as a problem for further research.

\end{document}